# Reversal of informational entropy and the acquisition of germ-like immortality by somatic cells

Marios Kyriazis

ABSTRACT

We live within an increasingly technological, information-laden environment for the first time in human evolution. This subjects us (and will continue to subject us in an accelerating fashion) to an unremitting exposure to 'meaningful information that requires action'. Directly dependent upon this new environment are novel evolutionary pressures, which can modify existing resource allocation mechanisms and may eventually favour the survival of somatic cells (particularly neurons) at the expense of germ line cells.

Here it is argued that persistent, structured information-sharing in both virtual and real domains, leads to increased biological complexity and functionality, which reflects upon human survival characteristics. Certain immortalisation mechanisms currently employed by germ cells may thus need to be downgraded in order to enable somatic cells to manage these new energy demands placed by our modern environment. Relevant concepts from a variety of disciplines such as the evolution of complex adaptive systems, information theory, digital hyper-connectivity, and cell immortalisation will be reviewed.

INTRODUCTION

Industrial and technology-driven developments have had a dramatic effect in lowering human mortality (Burger et al. 2012). Evolutionary natural selection pressures at maturity push mortality at the lowest possible level. Nevertheless, the lowest possible mortality rates achieved by natural evolution are significantly above the lowest levels achieved through technology in post-industrial countries. In countries such as Japan and Sweden the level is significantly lower (in some cases over 200-fold lower) compared to what natural selection has ever achieved. The reduction in human mortality rates has been achieved mostly during the past 100 years, through technology (i.e. structured and targeted information) with improved health care and socio-political risk-reduction. Based on this, it could be argued that the appropriate use of technology could further alter human mortality rates, not just through general health improvements and reduction of risks, but also by other mechanisms directly related to technology (such as global information sharing and processing). This environmentally-dependent malleability in mortality is inherent in the human genome and allows increases in life expectancy that are not necessarily based on genetic change (Burger et al. 2012).

Theories, hypotheses and ideas about the causes of age-related mortality are abundant (Viña et al. 2007). However, these contain limiting and implicit assumptions based on a 'component-approach':

that the components of the organism or cell are damaged during ageing, and that their repair may lead to rejuvenation. More recent efforts at understanding ageing are based instead upon a 'connection-approach': that the study of *how* individual agents (cells, macromolecules, organisms) are interconnected can provide a better perception of the complex mechanisms involved in ageing. The following discussion depends primarily on the 'connection-approach' and on our understanding of the opportunities afforded by strongly networked systems. It is specifically directed at studying the possible effects upon the maximum human lifespan. Four essential tenets in this discussion are:

1. **Information plus organization creates complexity, and complexity increases functionality** (Heylighen 2008). Increased functionality improves fitness, and this increases survival (FIGURE 1). Thus, in a specified niche, information increases survival (Coffrey 1998, Ben-Jacob et al. 2000) (TABLE 1). It was shown (Frieden and Gatenby, 2011) that a high-information, stable entropy state is associated with healthy living systems, whereas a phase transition to a low-information state is associated with cancer and aging.

2. **Nature tends, on the whole, to follow a 'from-simple-to-complex' paradigm** (William et al. 1990). There is a general propensity that leads to an increased complexity, self-organisation and dynamical variability (Furusawa and Keneko 2000, Adami et al. 2000). Simple initial events progress hierarchically to reach, through emergence, higher levels of complexity (Heylighen 1989). Higher complexity means increased functionality and increased survival (FIGURE 1).

3. **Ageing is accompanied by loss of information and complexity** i.e. increased entropy over time (Takahashi et al 2012, Lipsitz 2006, Cheng et al. 2009, Goldberger et al 2002, Lipsitz and Goldberger 1992). This loss is rooted in the suboptimal conditions caused by the uneven distribution of resources favouring the survival of the germ-line versus somatic repair (Kirkwood and Melov 2011). The rate of somatic repair tends to become progressively compromised as a function of age, resulting in accumulation of damaged biological material that reduces organisation and functionality i.e. a reduced information content, reduced complexity and thus reduced survival (Weon and Je 2010).

4. **A reversal of age-related information loss must lead to improved survival.** Based on the assumption that there is an adaptive value in the tendency to progress 'from simple to complex' it is reasonable to suggest that, if there is an artificial way to increase biological complexity, then this would have an impact upon somatic survival characteristics (Kyriazis 2003). Increased input of meaningful and actionable information (see Appendix) may be useful in restoring the information content of the organism (Kyriazis 2010) and thus it may herald a return to the 'simple to complex' progression model, with increased functionality, and thus increased survival (FIGURE 1).

This discussion will examine the details and mechanisms of premise number 4, namely that an appropriate increase of informational exposure (TABLE 2) may have a positive impact upon functionality, modulating energy allocation mechanisms which may shift more resources for somatic repair, thus resulting in somatic rejuvenation and increased survival of the individual.

FIGURE1

Information

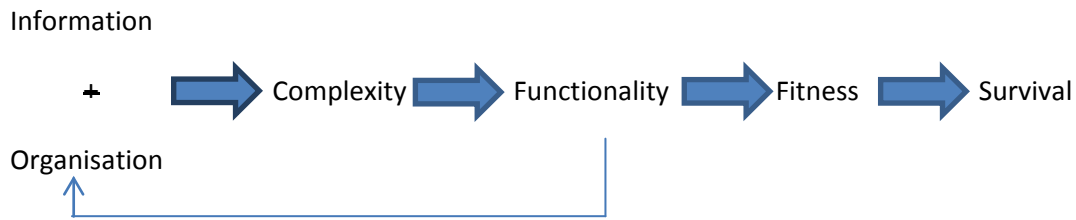

Organisation

Figure 1. Information (plus organisation) increases complexity and this increases functionality. This improves fitness and thus survival. (Functionality also increases organisation). Any increase in internal fitness requires the formation of new links and the strengthening of the interconnectedness between its nodes i.e. increased complexity, thus increased fitness and increased survival.

******************************************************************************

******************************************************************************

TABLE 1

Definitions:

Information = A meaningful set of data or patterns which influence the formation or transformation of other data or patterns, in order to reduce uncertainty and help achieve a goal

Organisation = The possession of a formal and co-ordinated structure

Complexity = Phenomena which emerge from a collection of functional interacting agents, with the potential to evolve over time

Functionality (biology) = The capacity to provide a useful function, leading to adaptive characteristics

Fitness = Resilience during change. The ability to survive and reproduce (although the significance of reproduction is drastically downgraded in this paper. The emphasis here is on survival rather than or reproduction).

******************************************************************************

DISCUSSION

It is well known that a challenging environment induces strategy adjustments that can up-regulate several parameters involved not only with brain function, such as BDNF (Brain Derived Neurotrophic Factor) (Vazquez-Sanroman et al. 2013), but also with age-associated immune impairment (Arranz et al. 2010), and other somatic benefits (Herring et al. 2010). An enriched environment (in the most general sense), is essentially a situation associated with increased exposure to actionable information, which will need to be assimilated and used in order to improve the function of the organism.

However, the long-range benefits of these improvements are not sufficiently known. While the health benefits of Environmental Enrichment (EE) have been well studied (for example: Kampermann et al. 2002) it is not clear what impact an enriched environment has on prolonging healthy lifespan, and certainly it is not clear if EE can prolong human lifespan well beyond the current maximum limit (approximately 122 years). Here, an attempt will be made to clarify some possible mechanisms involved in this respect, particularly with regards to examining whether EE (and thus information) can have any effects on prolonging human lifespan beyond the current maximum limits. With particular reference to the premise number 4 (above) the argument will examine the sequence of events between the detection and manipulation of information by neurons, the resources used for this process, and the impact this may have on the immortalisation mechanisms employed by germ cells.

It is necessary to describe in some detail the exact mechanisms that take place during the detection, acquisition, assimilation and elaboration of new information by the neuron. In order to generate signals and thus be able to control information, neurons use a substantial amount of energy. In an energy-efficient, low entropy organism the allocation of this energy must follow guidelines which are evolutionarily beneficial to the survival of the organism. Resources are finite and the organism must make arrangements for trade-offs (such as maintenance of the function of the germ line at the expense of somatic repair). However, the processes of handling a relentless arrival of actionable new information may eventually cause a shift from this scenario with remarkable consequences.

Using the example of a visual input which carries information that can be beneficial to the organism, we can study the sequence of events that are associated with the process. Photon catch and signal processing by retinal cells is dependent upon optimisation models that regulate parameters such as optical blur, random neural noise, and energy usage (Atick 1992). When new information is captured by retinal cells there is photonic recycling on the cell surface (Saxena et al. 2003). Ligand-receptor associations alter the conformation of the extracellular portion of intramembranous proteins and this change is transmitted to the cytoplasm by the trans-membranous helical segments by non-linear vibrations of proteins and generation of soliton waves (Saxena et al. 2003). In addition, photon-photon interactions induce molecular vibrations responsible for bio-amplification of weak signals described by:

$$mc^2 = BvLq$$

where *m* is the mass of the molecule, *c* is the velocity of the electromagnetic field, B is the magnetic flux density, v is the velocity of the carrier in which the particle exists, L is its dimension, and q is a unit charge. Following the successful activation of the retinal cell with the new information, there is generation of action potentials towards the visual cortex. For example, in the primary visual cortex V1, an early response consists of tiled sets of selective spatiotemporal filters. In the spatial domain, the functioning of V1 can be thought of as similar to many spatially local, complex Gabor transforms (Cope et al. 2009). The filters must be sufficiently effective to distinguish between useful (potentially relevant) and not useful (background noise) information, thus these need to be operating at an efficient level, more so when the influx of new information is sustained. Na+/K+ pumps concentrated in the axons and at synapses require significant amounts of energy in order to signal efficiently. Neurotransmitters and second messengers also add to the energy cost.

All of the processes mentioned in this example use energy, for example, through hydrolysation of ATP, which recycles signalling messengers and other molecules (Rando 1991). This energy is precious within an organism which allocates resources to the germ-line so efficiently. The supply of metabolic energy constrains the efficient transmission of information and repair of the neurons. However, if new information arrives at the sensory organ, it must be processed and thus the process will inevitably demand energy from any available resources. Within this framework, we can hypothesize that, if the amount of meaningful and actionable information is sufficiently large, and sufficiently sustained, then there will come a point when the ratio between the resources available to the germ line and resources available to the soma (for information processing AND subsequent somatic cell repair) will be forcibly shifted in favour of the latter. In other words, the information assimilation process may initiate a sequestration sequence that diverts resources from germ cells to somatic cells. It is important to emphasize that this is taken in the evolutionary context of increased fitness within a technological, information-laden niche. Thus, information processing carries an adaptive value within this specific niche.

While it is true that the energy used during signalling constrains the flow of information both within and between neurons, a sustained input of information makes such energy demands on the system that may eventually cause a transition to a new state whereby the energy available to the cell will be (in some way) proportionate to the amount of information which the cell is required to process. This energy may thus be used not only for information processing but, necessarily, also for cell repair in order to make information processing possible. It is unlikely that a high level of efficient information processing takes place in a damaged cell. The cell needs to be repaired (this includes age-related damage repair) in order to be in a position to process the information. Indeed, studies have estimated that 50-80% of energy consumption is allocated to signalling, and the remainder is allocated to cell maintenance and repairs (Laughlin 2001). There is evidence to show that the organism tries to economise in energy expenditure by minimising the costs associated with signal processing, such as minimising the signal-to-noise ratio and band-width (Anderson and Laughlin 2000). However, a cost-efficient neural system increases the numbers of active cells and reduces waste by reducing inactive neurons. Here a balance must be found between the energy costs of information processing and the contribution to overall fitness made by this information. If the fitness of the organism increases as a result of the new information, then it means that the cost of processing that information was worth it. This will have a secondary effect on other associated processes such as metabolic, immune, vascular and respiratory mechanisms, all of which must also be repaired. It must be reiterated that this discussion refers specifically to the evolutionary fitness of modern humans who live in an increasingly technological niche, where survival depends on adapting successfully to, ultimately, relentless information processing.

The Role of Phase Transition

Recent developments in our understanding of phase transitions provide useful insights applicable to this discussion. A phase transition is a profound structural re-organisation when there is a large change in resource availability for maintenance. Sudden, rapid-acceleration phase transitions are associated with emergence, mutations, breakthroughs and development of autocatalytic systems which change the *status quo* and result in new situations which may improve fitness and survival (de Rosnay 2000).The study of such threshold dynamics gives interesting insights into plausible mechanisms for artificial manipulation of the process (Paperin et al. 2011). Kiss et al. (2009) suggest that in order for a phase transition to take place it is necessary to have increased complexity **and** increased exposure to new challenges **and** a reduction in available resources. In the case under consideration here, the continuous exposure to new actionable information, and the unrelenting energy demands associated with it, may lead to a critical point where the value of selective pressure induces a phase transition in certain basic biological parameters.

Theoretically, it was shown that any continual additions to a set of agents lead to a critical transition from disconnected to connected dynamics. This phase transition depends not only on the quality and quantity of the added agent but also on critical thresholds within the set of agents (Erdös and Rényi 1960). To mention a biological analogy, it was shown that sensory stimuli trigger rapid and chaotic firing of neurons, which is fed through feedback loops and gradually re-synchronises neural firing. Cognition involves sudden transitions in cortical electroencephalography oscillations, and suggests that there is a dependence on connectivity phase transitions (Freeman 1975).

The suggestion that a sustained exposure to actionable information forcibly induces a change in basic hitherto stable evolutionary processes is not an entirely speculative one. It has been proposed that increased aerobic activity (thus in this case physical, instead of cognitive, effort) had a direct evolutionary impact on the human brain (Raichlen and Polk 2013). The increased and sustained aerobic activity of hunter-gatherers resulted in the up-regulation of peripheral BDNF and increased its concentration to such a degree that it eventually crossed the blood-brain barrier and directly influenced the function and survival of the neurons. The plasma concentrations of other factors that are normally found in the periphery (i.e. not in the brain) such as insulin-like growth factor 1 (IGF-1) and vascular endothelial growth factor (VEGF) all of which increase during physical activity, were elevated by the sustained physical effort and, having crossed the blood-brain barrier, reached the brain resulting in improved neurogenesis. Thus, it is hypothesized that other sustained activities, in this case cognitive stimulation, may have physical effects which cause a phase transition from one evolutionary stage to another.

It was shown that, by increasing transcriptional noise (through the manipulation of free energy availability, and thus the selective pressure) it may be possible to trigger a structural phase transition where genetic networks change their topology from random to segregated (Peixoto 2012). In this case, most genes are regulated by a smaller and denser subset of genes (i.e. by transcription factor genes that regulate all others). This could be a mechanism used in germ line immortalisation and it could possibly also be made to operate in putative somatic cell immortalisation, through manipulation of the parameters. The process is described by the order parameter:

$$\phi = \frac{b* - br}{bmin - br}$$

where *br* is the value of *b\** -the average error (the main fitness criterion) for a fully random network, and *bmin* is the smallest possible value of *b\** for a given *k* (the average number of inputs per node).

*bmin* is given by $\sum_k pk\ mk\ (p)$

(*m* denotes any cell from a system of many cells)

When ∅=0 the network is fully random (i.e. unfit), whereas when ∅=1 the network has the largest possible value of fitness. The process therefore is dependent upon the number of inputs per node (*k*), and the noise level *p*, that is to say, it is dependent upon the overall degree of information input, which may modify selection pressure in a way favourable to somatic cells. Theoretical studies of evolutionary phase transitions in complex adaptive systems (CAS) (Scott 2013) show that, when agents compete for limited resources, they tend to self-segregate into opposing groups depending on the 'prize-to-fine' ratio (i.e. the probability of fitness success vs fitness failure).The evolutionary kinetics of both germ-cells and somatic cells (both are CAS) can then be studied using the above model. Both groups of cells can be considered as agents (of systems) competing for resources and are thus subjected to evolutionary dynamics rules. It was shown that these systems may change their characteristics from 'cautious' agents to 'extreme' ones and back (i.e. from random to segregated, via a phase transition as described above) depending on the values of the prize-to-fine ratio, i.e. when the value of resources available for maintenance changes (Xu et al. 2005. This may explain the mechanisms of the phase transition which depend on resource distribution: when the 'prize' of following a certain decision to allocate resources and repair something is higher than the 'fine' associated with that repair [i.e. within a defined fitness landscape, when the fitness increase following a particular repair is more than the fitness reduction associate with that repair], then there will be a transition that favours that particular repair which carries a better survival value. Of course, a better survival value is currently achieved through favourable repairs to germ line cells at the expense of somatic cells. However, it is suggested that this balance may be shifted at the favour of somatic **if** the prize to fine ration is changed, via increased demand for resources following informational exposure (thus via manipulating the number of inputs per node(cell).

Germ-line Immortality and Possible Reversal Mechanisms

Germ cells achieve immortality by ensuring that they maintain robustness - the redundancy that counteracts the effects of random noise. Evolution drives the balance of the right trade-off between robustness and maintenance resources. The trade-off between survival of the somatic cells and reproduction could be due to factors such as:

a. The impossibility to maintain all processes within the body indefinitely, due to lack of resources. Or
b. The damage to the somatic cells could be incurred during the normal course of reproduction (Partridge et al 2005).

There is a direct relationship between increasing age and genomic instability in somatic cells. Maintenance resource reallocation favours reproduction at the expense of somatic cell senescence. However, it is possible to encounter soma-to-germ line transformation of gene expression which is normally encountered only in germ-line. It was shown (Curran et al 2009) that somatic cells of insulin-like signalling mutants are efficiently protected against genotoxic stress and exhibit more germ-line features. Therefore, the possible germ line-like (and thus immortality-affording) transformation of somatic cells is not*, a priori*, excluded.

The rejuvenation process encountered in germ line erases the age-related damage that accumulates over the years (Mendvedev 1981). Three mechanisms have been suggested to account for germ line immortality (Smellick and Ahmed 2005):

1. A generally more efficient and increased rate of cellular repair and maintenance, as well as specific repair and rejuvenation mechanisms.

2. Efficient selection of fully functional germ cells which are allowed to propagate (Murphey et al 2013) at the expense of less efficient germ cells. A relevant concept here is that of apoptosis regulation.

3. Non-autonomous contribution of the soma to germ line immortality (Kirkwood 1987)

Such strategies may also be present in somatic cells, but are significantly down-regulated (Smellick and Ahmed 2005). The aim here is to examine the possibility that this rejuvenation process can be driven to operate in somatic cells (Avise 1993). It is important to highlight that certain mechanisms of germ line rejuvenation could be dependent upon epigenetic modifications and factors that regulate transcription (Santos and Dean 2004).Thus it is also conceivable that careful epigenetic co-ordination may have certain rejuvenating effects upon somatic cells.

Mechanism number 3 above, is particularly attractive as it raises the possibility that somatic cells may, under certain circumstances, withhold those immortality contributions for their own repair (when for example, somatic cells are under intense pressure to maintain themselves). It is known that ablation of germ line extends lifespan in drosophila and in C. elegans, suggesting that defects in germ line immortality may provide beneficial resources to the soma (Arantes-Oliveira et al 2002).

Fontana (2010) has suggested that when 'good' (i.e. immortality-affording) matching sequences are not present in the germ-line genome, then these 'good' sequences are created in the somatic cell and then migrate to the germ line through the bloodstream (Germ Line Penetration). In this case, it has been suggested that transposition of genomic elements in somatic cells drives differentiation in germ cells that drives evolution. As a consequence, it can also be argued that this process can conceivably be arrested, with somatic cells retaining the immortality-affording sequences and using these for their own repairs. Recent finding that neurons exhibit quasi-immortality characteristics (i.e. they are able to live longer than their host), may suggest that somatic cells that are evolutionarily beneficial are retained and maintained, perhaps indefinitely (Magrassia et al. 2013).

CONCLUDING DISCUSSION

Kirkwood (2008) describes menopause as the result of a convergence of biological and cultural evolution. Menopause has evolved as an adaptation to the biological value it has conferred upon older women. This implies, as a general concept, that if a situation confers sufficient fitness to a group of people then it will evolve and be favourably selected. Therefore, the same principle may be operating in the situation discussed here: that if the increased and continual input of information confers an increased fitness to individuals (within a niche of information technology) then the associated processes will be favourably selected and evolve. It is known that epigenetic regulation of gene expression responds to external environmental factors, such as increase environmental enrichment, a fact that provides a genetic basis for the improvement of the functional ability of the individual (Vaiserman 2011). An example of how information-enriched environments may impact on evolution is the observed increased overall intelligence is recent times. Flynn (1987) has shown that in the developed countries the average Intelligence Quotient has been rising by about 3 points per decade. A probable explanation for this is that higher intelligence is the result of environmental (i.e. artificial, information-laden) factors such as improved medical care and general health, better education and higher cognitive stimulation by living in an increasingly complex society (Heylighen 2010). This provides a direct link between increased meaningful information load and a resulting biological benefit.

Evolution depends on innovations in energy extraction mechanisms. Yun et al. (2006) have suggested that energy efficiency is a driving force in evolution as it increases fitness, and that the 'variation-selection' method of Darwinian evolution may not be an all-encompassing biological paradigm. They suggest that a model of evolution based on the 'concept of competition for energy' *"…encompasses prior notions of evolution and portends post-Darwinian mechanisms… Under these circumstances, indefinite persistence may become favored over life-death cycling, as usage of energy may then occur more efficiently within a single lifespan rather than over multiple generations"*. If 'competition for energy' is a valid evolutionary model, then the possibility arises that increased competition for available energy resources may reverse the current supremacy of virtually immortal germ cells at the expense of somatic ones. This competition for resources may be intentionally intensified through strategies described in TABLE 2. These artificial strategies may have a deep impact upon our biological processes with regards to resource allocation.

*****************************************************************************

TABLE 2

Strategies for enhancing information exposure

The following are some examples of how to increase input of information that requires action. The information encourages a response and facilitates the propagation of that response, with biological repercussions both within the brain and in the body as a whole (for quantification, see Appendix).

1. Enriched environment (Kemermann et al 2002), hormetic challenges (Kyriazis 2010).

2. Social hyper-connectivity (real and virtual/online) (Kyriazis 2012).

3. Behavioural models such as a goal-seeking behaviour (Wilson 2000), search for excellence, and a bias for action (Peters and Waterman 2004).

4. The pursuit of innovation, diversification, creativity, (Mumford et al. 1988), novelty (Strange et al. 2005), and the avoidance of routine.

*****************************************************************************

Within hierarchical networks, the element of co-operation is essential. It was shown that this co-operation can transpose fitness from a simpler level to a higher level, and in doing so it extends the fitness and survival of the network (Michod 2006). In the specific example of human survival it may be argued that increased cooperation and network complexity will transcend the necessity of fitness of the germ line (germ cells and DNA) and increase, instead, the fitness of the more complex network (i.e. the brain, and as a consequence the entire organism).  These more complex, globally co-operating networks are ideal solutions within our modern, rapidly changing technology-driven environment and they provide a situation whereby damage control, maintenance and restructuring may result in delayed ageing (i.e. delayed loss of complexity and functionality) (Kiss et al 2009). These authors have also suggested that in order to preserve the evolvability of the network it is necessary to maintain a pool of creative elements within the network, such as  molecular chaperones (in cellular networks), stem cells (in complex organisms) and creative persons (in social networks). Thus, the evolutionary need for maintaining the soma at the expense of the germ line becomes evident across all platforms and scales.

APPENDIX

Quantifying informational input

During everyday life we are exposed to **random** unintentional challenges and information (cognitive challenges, novelty of the environment, new ideas and situations). This helps maintain brain function. At the same time, we experience increasing entropy, which eventually causes death because the degree of information input tends to zero with time, whereas entropy tends to infinity, therefore Ai =0 (death) – see below.

The proposal of this discussion is to introduce another variable, namely the sum of **intentional** exposure to information that requires action (i.e. intentional stimulation, engagement and integration with digital communication, goal-oriented behaviour, novelty-seeking etc. see TABLE 2) which adds robustness into the equation, delaying the Ai=0 scenario (as below).

The equation is as follows:

$$Ai = \frac{\sum_{k=1}^{n} \alpha k + \sum_{k=1}^{n} \beta k}{\Delta s} - \Delta l$$

*Ai* is the degree of activation of the system (in this case, the system is the brain)

*αk* is the sum of all individual **random** exposures to cognitive stimulation during everyday life *(α1+α2+α3+ …αn)*

$\sum_{k=1}^{n} \alpha k$ is the total sum of all such sums of exposures where *k* ranges from 1 to *n*

*βk* is the sum of all individual **intentional** exposures to cognitive stimulation *(β1+β2+β3+…βn)*

$\sum_{k=1}^{n} \beta k$ is the total sum of exposures where *k* ranges from 1 to *n*

$\Delta s$ is total entropy of the system

$\Delta l$ is a variable to account for the degree of individual loss of intelligence as global intelligence increases

*Ai*=0 is death (infinite entropy and zero information)

*Ai=k* , is ideal, healthy (non-ageing) status

*Ai> k,* is over-stimulation (unhealthy)

*Ai< k,* is under-stimulation (unhealthy)

***************************************************************************